\documentclass[a4paper,12pt]{article}
\usepackage{graphicx,amssymb,bm,latexsym}

\begin{document}
\topmargin 0pt \oddsidemargin 0mm
\renewcommand{\thefootnote}{\fnsymbol{footnote}}
\begin{titlepage}
\vspace{5mm}
\begin{center}
{\Large \bf Note on I-brane Near Horizon PP-wave Background}
\vskip 15mm
{Rashmi R. Nayak $^a$ \footnote{Email: rashmi@cts.iitkgp.ernet.in},
Pratap K. Swain$^b$\footnote{E-mail: pratap@phy.iitkgp.ernet.in}}
\vskip 10mm
{\it $^a$ Centre for Theoretical Studies \\
Indian Institute of Techology Kharagpur, Kharagpur-721 302, India }
\vskip 10mm
{\it $^b$ Department of Physics and Meteorology\\
Indian Institute of Technology Kharagpur, Kharagpur 721302, India}

\end{center}
\vspace{15mm} \centerline{{\bf{Abstract}}} \vspace{5mm}
We find out a PP-wave spacetime with constant Neveu-Schwarz (NS) three form flux by taking a
Penrose limit on the 1+1 dimensional intersection of two orthogonal stacks of
fivebranes in type IIB string theory. We further find out an intersecting (D1-D5)-brane solution in
this background and analyze its supersymmetry properties by solving the dilatino and
gravitino variations explicitly.
\begin{flushleft}
\end{flushleft}
\end{titlepage}

\section{Introduction}
Study of intersecting brane solutions in supergravities have played an important role in
understanding the underlying gauge theories and their dynamics. The so called I-brane
background which arises in the 1+1 dimensional intersection of two orthogonal stacks of
five branes in type IIB string theory has been one of the interesting exact string background
that has drawn lot of attention in recent past. On one hand from the holography arguments
one could expect that the dynamics at the intersection
should be holographically related to a 2+1 dimensional bulk theory, on the other hand there
is an enhancement of symmetry in the near horizon geometry. Naively,
looking at the theory at the intersection one can expect that this theory should be
invariant under (1+1)-dimensional Poincare symmetry $ISO(1,1)$, however,
as derived in \cite{Itzhaki:2005tu, Lin:2005nh} the near horizon geometry describes a
2 + 1 dimensional theory with Poincare symmetry $ISO(2,1)$, and twice as many supercharges one
might expect. Furthermore, in \cite{Kluson:2005eb}, these interesting properties were studied from the point
of view of D1-brane probe. It was shown that the enhancement of the near horizon
geometry has clear impact on the worldvolume dynamics of the D1-brane probe.
Further analysis on these and related topics have been discussed in \cite{Kluson:2005qq, Hung:2006jh,Hung:2006nn,Kluson:2007st}.

In the present note we would like to take a Penrose limit on this geometry and look for
exact string background. Penrose limit on supergravity D-brane backgrounds has been
considered in, for example \cite{Blau:2002mw}. Further Penrose limits on backgrounds
that correspond to non-local theories
has been studied earlier (see for example \cite{Hubeny:2002vf, Alishahiha:2002zu}). For example the Penrose limit on the near horizon geometry of
a stack of NS5- branes, was shown to produce a worldsheet theory which is free. Though the
corresponding PP-wave background was shown to be supersymmetric, the D-branes
seem to break all spacetime superymmetries although they preserve the same worldsheet superymmetries
as that of flat space.
This issue was resolved in \cite{Hassan:2003ec} by showing that these
supersymmetries do not have zero modes on the worldsheet and hence do not
admit local space-time realizations. Hence, unlike the perturbative
spectrum, the D-brane spectrum of strings in the linear-dilaton pp-wave is not similar to the flat space case. We would like to further anlayze the fate of these facts in the case of PP-wave background
coming from two stacks of intersecting NS5- branes. We find that the Penrose limit on the I-brane
geometry gives a background which preserves 1/2 supersymmetries although the parent
solution preserves 1/4 spacetime supersymmetries.  Further, for certain values of
parameters this reduces to the well known $H_6$ PP-wave background with
constant NS-NS 3-form flux.
So the sigma model action can be written down directly by following \cite{Russo:2002rq}, and one can also study classical D-brane solutions in this background.
We further present an
example of (D1-D5)-brane solution in this background and analyze the spacetime superymmetries.
The rest of the paper is organized as follows. In section-2 we find out a PP-wave background
by taking a particular type of Penrose limit on the I-brane background and find that it preserves half
of the total spacetime superymmetry. In section-3, we write down the supergravity solution of a class of D1-D5 solution and analyze its supersymmetries. Finally in section-4, we conclude with some comments.

\section{Penrose limit of I-brane background}
We consider the intersection of two stack of NS5-branes on $R^{1,1}$ \cite{Itzhaki:2005tu}. We
have $ k_1 $ number of NS5- branes extended along $(0,1,2,3,4,5)$
directions and $k_2$ number of NS5- branes extended along $(0,1,6,7,8,9)$ directions.  For
writing down the explicit supergravity background corresponding to the above configuration, let us define
\begin{equation}
\textbf{y} = (x^2,x^3,x^4,x^5),\>\>\>
 \textbf{z} = (x^6,x^7,x^8,x^9).
\end{equation}
Further assume that $k_1$ NS5- branes are localized at the points $z_n, n = 1,...,
k_1$, and $k_2$ NS5-branes are localized at the points $y_a$, $a = 1,...,
k_2$. Every pairs of fivebranes from different sets intersect at
different point $(y_a,z_n)$. The supergravity background
corresponding to this configuration takes the form\cite{Itzhaki:2005tu}
\begin{eqnarray}
\Phi(z,y) = \Phi_1(z) + \Phi_2(y),\nonumber \\
g_{\mu\nu} = \eta_{\mu\nu},  \>\>\> g_{\alpha\beta} = e^{2(\phi_2 - \phi_2(\infty))}
\delta_{\alpha\beta}, \>\>\>  g_{pq} = e^{2(\phi_1 - \phi_1(\infty))}
\delta_{pq}, \nonumber \\
H_{\alpha\beta\gamma} = -\epsilon_{\alpha\beta
\gamma\delta}\partial^{\delta}\Phi_2, \>\>\> H_{pqr} = -\epsilon_{pqrs} \partial^{s}\Phi_1,
\label{I-bkg}
\end{eqnarray}
where $(\mu,\nu = 0,1), (\alpha,\beta, \gamma,\delta = 2,3,4,5)$, and $(p,q,r,s = 6,7,8,9)$,
and $\Phi$ on the first line of eqn. (\ref{I-bkg}) is the dilaton defined as
\begin{eqnarray}
 e^{2(\Phi_1 - \Phi_1(\infty))} &=& 1 + \sum_{n=1}^{k_1}\frac{{l_s}^2}{(z-z_n)^2},
\cr & \cr e^{2(\Phi_2 - \Phi_2(\infty))} &=& 1 + \sum_{a=1}^{k_2}\frac{{l_s}^2}{(y-y_a)^2}.
\end{eqnarray}
Our goal is to find a PP-wave background of this geometry when
$z_n = y_a = 0$. To simplify our notation let us denote
\begin{equation}
 e^{2(\Phi_1 - \Phi_1(\infty))} = H_1(z), \>\>   e^{2(\Phi_2 - \Phi_2(\infty))} = H_2(y),
\end{equation}
where for coincident branes the harmonic functions are given by
\begin{equation}
 H_1 = 1 + \frac{k_1{l_s}^2}{z^2}, \>\>   H_2 = 1+ \frac{k_2{l_s}^2}{y^2}.
\end{equation}
Let us now consider the the near horizon geometry of the above configuration, i.e.
\begin{equation}
 \frac{k_1{l_s}^2}{z^2} >> 1, \>\>   \frac{k_2{l_s}^2}{y^2} >> 1
\end{equation}
so that the harmonic functions take the following form
\begin{equation}
 H_1 = \frac{\lambda_1}{{r_1}^2}, \>\>  \lambda_1 = k_1{l_s}^2, \>\>  H_2 = \frac{\lambda_2}{{r_2}^2}, \>\>   \lambda_2 = k_2{l_s}^2.
\end{equation}
Then the metric in eqn. (\ref{I-bkg}) takes the form
\begin{equation}
ds^2 = -dt^2 + (dx^1)^2 + \frac{\lambda_1}{{r_1}^2}d{r_1}^2 +
{\frac{\lambda_2}{{r_2}^2}}d{r_2}^2 + \lambda_1d{\Omega_1}^{(3)} +
\lambda_2d{\Omega_2}^{(3)},
\label{I-bkg1}
\end{equation}
where $d{\Omega_1}^{(3)}$ and  $d{\Omega_2}^{(3)}$ correspond to the
line elements on the unit sphere along (2,3,4,5) and (6,7,8,9) directions respectively. To describe them further we introduce the following coordinates
\begin{eqnarray}
x^2 + ix^3 &=& r_1\cos\theta_1e^{i\phi_1}, \>\>  x^4 + ix^5 = r_1\cos\theta_1e^{i\psi_1}, \nonumber \\
x^6 + ix^7 &=& r_2\cos\theta_2e^{i\phi_2}, \>\>  x^8 + ix^9 = r_2\cos\theta_2e^{i\psi_2},
\end{eqnarray}
so that the volume elements on the sphere, and the 2-form potentials which give the
volume forms in eqn. (\ref{I-bkg}) are given by
\begin{eqnarray}
d{\Omega_1}^{(3)} &=& d{\theta_1}^2 + \sin^2\theta_1d{\phi_1}^2 +
\cos^2\theta_1d{\psi_1}^2, \nonumber \\
b_{\phi_1\psi_1} &=& \lambda_1\cos^2\theta_1, \>\> 0 < \theta_1 < \frac{\pi}{2}, \>\>
0 = \phi_1,\psi_1 < 2\pi, \nonumber \\
d{\Omega_2}^{(3)} &=& d{\theta_2}^2 + \sin^2\theta_2d{\phi_2}^2 +
\cos^2\theta_2d{\psi_2}^2, \nonumber \\
b_{\phi_2\psi_2} &=& \lambda_2\cos^2\theta_2, \>\> 0 < \theta_2 < \frac{\pi}{2}, \>\>
0 = \phi_2,\psi_2 < 2\pi,
\end{eqnarray}
To proceed further, as a first step let us introduce two modes $\rho_1$ and
$\rho_2$ defined by
\begin{equation}
r_1 = e^{\frac{\rho_1}{\sqrt{\lambda_1}}}, \>\>
r_2 = e^{\frac{\rho_2}{\sqrt{\lambda_2}}} \ ,
\end{equation}
and introduce two modes $r, y$ through
\begin{equation}
Qr = \frac{\rho_1}{\sqrt{\lambda_1}} +
\frac{\rho_2}{\sqrt{\lambda_2}}, \>\>
Qy = \frac{\rho_1}{\sqrt{\lambda_2}} -
\frac{\rho_2}{\sqrt{\lambda_1}},
\label{12}
\end{equation}
where
\begin{equation}
Q = \frac{1}{\sqrt\lambda}, \>\>  \frac{1}{\lambda} =
\frac{1}{\lambda_1} + \frac{1}{\lambda_2}.
\end{equation}
Note that the inverse transformations of (\ref{12}) take the form
\begin{eqnarray}
\rho_1 &=& \frac{1}{\sqrt{\lambda_1 + \lambda_2}}\big(
\sqrt{\lambda_1}y + \sqrt{\lambda_2}r\big),
\cr & \cr \rho_2 &=& \frac{1}{\sqrt{\lambda_1 + \lambda_2}}\big(
\sqrt{\lambda_1}r - \sqrt{\lambda_2}y\big).
\end{eqnarray}
Note that this transformation implies that the dilaton
is a function of $r$ only, i.e.
\begin{eqnarray}
\Phi &=& \Phi_1 + \Phi_2 = \frac{1}{2}(H_1 + H_2) +
\Phi_1(\infty) + \Phi_2(\infty)
\cr & \cr &=& -\frac{\rho_1}{\sqrt{\lambda_1}} -
\frac{\rho_2}{\sqrt{\lambda_2}} + \Phi_0 = -Qr + \Phi_0.
\end{eqnarray}
After doing this substitution, the metric in eqn. (\ref{I-bkg1}) and the
NS 2-form potentials take the following form
\begin{eqnarray}
ds^2 &=& -dt^2 + (dx^1)^2 + dr^2 + dy^2 +
 \lambda_1d{\Omega_1}^{(3)} + \lambda_2d{\Omega_2}^{(3)} \ , \nonumber \\
b_{\phi_1\psi_1} &=& \lambda_1\cos^2\theta_1, \>\>\>
b_{\phi_2\psi_2} = \lambda_2\cos^2\theta_2.
\label{I-bkg2}
\end{eqnarray}
Now we are ready to take a Penrose limit on this geometry. For the time
being let us keep $\lambda_1 = \lambda_2 = N^2$.
One can think that this is an example of asymmetric pp-wave coming from the intersection of two stacks on NS5-branes on a string. Now rescaling $dt = N^2 dt$, the metric and the NS 2-form potentials become,
\begin{eqnarray}
ds^2 &=& N^2\bigg(-dt^2 + \cos^2\theta_1d\psi_1^2 +
d\theta_1^2 + \sin^2\theta_1d\phi_1^2 + \cos^2\theta_2d\psi_2^2 + d\theta_2^2 +
\sin^2\theta_2d\phi_2^2\bigg) \nonumber \\
&+&  (dx^1)^2 + dr^2 + dy^2 \nonumber \\
b_{\phi_1\psi_1} &=& N^2 \cos^2\theta_1, \>\>  b_{\phi_2\psi_2} = N^2\cos^2\theta_2.
\end{eqnarray}
Let us make the following substitution to mix coordinates among the two
spheres
\begin{eqnarray}
\psi_1 = \cos\alpha\psi_1 + \sin\alpha\psi_2, \>\> \psi_2 &=& -\sin\alpha\psi_1 + \cos\alpha\psi_2 .
\end{eqnarray}
Then the metric becomes
\begin{eqnarray}
ds^2 &=& N^2\bigg(-dt^2 + (\cos^2\theta_1
\cos^2\alpha + \cos^2\theta_2\sin^2\alpha)d\psi_1^2
+ (\cos^2\theta_1\sin^2\alpha + \cos^2\theta_2\cos^2\alpha)d\psi_2^2 \nonumber \\
&+& d\theta_1^2 + \sin^2\theta_1d\phi_1^2 + d\theta_2^2 +
\sin^2\theta_2d\phi_2^2 + 2\sin\alpha\cos\alpha(\cos^2\theta_1 - cos^2
\theta_2)d\psi_1d\psi_2\bigg) \nonumber \\
&+& (dx^1)^2 + dy^2 + dr^2. \nonumber \\
\end{eqnarray}
Lets us now take the Penrose limits as follows \cite{Lu:2002kw}:
\begin{eqnarray}
t &\rightarrow& x^+ + \frac{x^-}{N^2}
\cr & \cr \psi_1 &\rightarrow& \frac{1}{\sqrt2}
\left(x^+ - \frac{x^-}{N^2}\right)
\cr & \cr \psi_2 &\rightarrow& \frac{\psi_2}{N},
\theta_1 \rightarrow \frac{\theta_1}{N},
\theta_2 \rightarrow \frac{\theta_2}{N}.
\end{eqnarray}
and take $N \rightarrow \infty$, the metric becomes
\begin{eqnarray}
ds^2 &=& -\frac{1}{2}(\theta_1^2\cos^2\alpha +
\theta_2^2\sin^2\alpha)(dx^+)^2 - 2dx^+dx^-
\cr & \cr &+&  d\psi_2^2 + d\theta_1^2 + \theta_1d\phi_1^2
 + d\theta_2^2 + \theta_2d\phi_2^2 + (dx^1)^2 + dy^2 + dr^2.
\end{eqnarray}
Note that the dilaton vanishes during this limit.
Further let us make the following substitutions
\begin{eqnarray}
z_1 &=& \theta_1\cos\phi_1,\>\> z_2 = \theta_1\sin\phi_1,\>\>\> z_3 = \theta_2\cos\phi_2,\>\>\>
z_4 = \theta_2\sin\phi_2 .
\end{eqnarray}
Now the metric takes the following form (after scaling $x^+ = 2\mu x^+$, and $x^- = - x^- / {2\mu}$, where $\mu$ is a constant)
\begin{eqnarray}
ds^2 = 2dx^+dx^- -2\mu^2[(z_1^2+z_2^2)\cos^2\alpha +
(z_3^2+z_4^2)\sin^2\alpha](dx^+)^2 + \sum_{i=1}^4dz_i^2 + \sum_{a=5}^8 dz_a^2.
\label{I-ppwaveg}
\end{eqnarray}
At the same time the NS-NS B-fields become
\begin{eqnarray}
B_{12} = 2\sqrt{2}\mu x^+ \cos \alpha, \>\>\>
B_{34} = 2\sqrt{2}\mu x^+ \sin \alpha \ .
\end{eqnarray}
which gives the constant 3-form NS-NS flux
\begin{eqnarray}
H_{+12} = 2\sqrt{2}\mu \cos \alpha, \>\>\>
H_{+34} = 2\sqrt{2}\mu \sin \alpha \ .
\label{I-ppwaveH}
\end{eqnarray}
Let us discuss this background in a bit more detail. Now for $\sin \alpha = \cos \alpha = \frac{1}{\sqrt{2}}$,
this is the well known $H_6$ PP-wave background coming from the $AdS_3 \times S^3$ geometry.
For $\alpha = 0, ~({\rm or}~ \pi/2)$, this is the background obtained by taking the near horizon pp-wave
limit of a stack of NS5-brane \cite{Hubeny:2002vf}. Let us now look at the supersymmetry this background preserve. It is
well known that the I-brane background preserves 1/4 superymmetries. One can check that this
background above preserves 1/2 of the total spacetime superymmetry after imposing the condition
\begin{eqnarray}
\Gamma^{\hat +}\epsilon_{\pm} = 0 \ .
\end{eqnarray}
The sigma model action can be written down directly following \cite{Russo:2002rq}.
Few comments are in order now. In fact one can start with a background which is the $S$-dual of
the one that is considered here, namely two stacks of $D5$-branes intersecting over a line and
take a Penrose limit. One will get the same background metric as above but the NS-NS flux will be replaced
by the Ramond-Ramond 3-form flux.  Furthermore, one can also uplift this to the $M5-M5'$
intersecting over a
2-plane and take the Penrose limit of the geometry. The $R-R$ three form flux  will be replaced by $R-R$ four form flux. Similar solutions are discussed in \cite{Cvetic:2002si,Cvetic:2002hi} earlier.

In fact one could further notice that this background with a light like linear dilaton is also a solution
of type IIB supergravity equations of motion. The metric, null dilaton and the 3-from flux in this case is given by
\begin{eqnarray}
&&ds^2 = 2dx^+dx^- -2\mu^2 (x^+)[(z_1^2+z_2^2)\cos^2\alpha +
(z_3^2+z_4^2)\sin^2\alpha](dx^+)^2 +  + \sum_{a=1}^8 dz_a^2, \nonumber \\
&&H_{+12} = 2\sqrt{2}\mu(x^+) \cos \alpha, \>\>\> H_{+34} = 2\sqrt{2}\mu (x^+) \sin \alpha \ , \Phi = \Phi(x^+) \nonumber \\
\end{eqnarray}
\section{Intersecting D-brane solutions}
We now write down the supergravity solution for intersecting (D1-D5)-branes in the background
(\ref{I-ppwaveg}) and (\ref{I-ppwaveH}).
The metric, dilaton and the field strengths
of such a configuration is give by
\begin{eqnarray}
ds^2 &=& (f_1 f_5 )^{-1/2} \left(2dx^+dx^- -2\mu^2 [(z_1^2+z_2^2)\cos^2\alpha +
(z_3^2+z_4^2)\sin^2\alpha](dx^+)^2\right) \nonumber \\
&+& \left(\frac{f_1}{f_5}\right)^{1/2}\sum_{a=5}^8 dz_a^2 + \left(f_1f_5\right)^{1/2}\sum_{i=1}^4dz_i^2, \>\>\> e^{2\phi} = \left(\frac{f_1}{f_5}\right)^{1/2} \nonumber \\
H_{+12} &=& 2\sqrt{2}\mu \cos \alpha, \>\>\> H_{+34} = 2\sqrt{2}\mu \sin \alpha \ , \nonumber \\
F_{+-i} &=& \partial_i f^{-1}_1, \>\>\> F_{ijk} =
\epsilon_{ijkl}\partial_l f_5,\>\>\> f_{1,5} = 1+ \frac{Q_{1,5}}{r^2} \ ,
\label{d1-d5}
\end{eqnarray}
where $f_{1,5}$ are the harmonic functions of the D1 and D5-brane in the transverse 4-space
respectively.
The above (D1-D5)-brane configuration solves all type IIB field equations. One can notice that for
$\cos \alpha = \sin \alpha =\frac{1}{\sqrt{2}}$, the above solution reduces to a (D1-D5)-brane
solution in a PP-wave
background coming from pure $AdS_3 \times S^3$ with NS-NS flux \cite{Biswas:2002yz}.
Further intersecting brane solutions can be found out by applying $T$-dualities along the
$z^{a}$ directions. One can further check that by putting $f_1 =1$, the above solution
reduces to a D5-brane lying along $(+,-,5,6,7,8)$ directions and by putting
$f_5 =1$, this reduces to a localized D1-brane lying along $(+,-)$ directions.

Now let us check the amount of unbroken supersymmetries the above (D1-D5)-brane solution
preserves. The supersymmetry variations of dilatino
and gravitino fields of type IIB supergravity in the string frame are
given by \cite{{Schwarz:1983qr},{Hassan:1999bv}},
\begin{eqnarray}
\delta \lambda_{\pm} &=& {1\over2}(\Gamma^{\mu}\partial_{\mu}\phi \mp
{1\over 12} \Gamma^{\mu \nu \rho}H_{\mu \nu \rho})\epsilon_{\pm} + {1\over
  2}e^{\phi}(\pm \Gamma^{\mu}F^{(1)}_{\mu} + {1\over 12} \Gamma^{\mu \nu
  \rho}F^{(3)}_{\mu \nu \rho})\epsilon_{\mp}\,,\\
\label{dilatino}
\delta {\Psi^{\pm}_{\mu}} &=& \Big[\partial_{\mu} + {1\over 4}(w_{\mu
  \hat A \hat B} \mp {1\over 2} H_{\mu \hat{A}
  \hat{B}})\Gamma^{\hat{A}\hat{B}}\Big]\epsilon_{\pm} \cr
& \cr
&+& {1\over 8}e^{\phi}\Big[\mp \Gamma^{\lambda}F^{(1)}_{\lambda} - {1\over 3!}
\Gamma^{\lambda \nu \rho}F^{(3)}_{\lambda \nu \rho} \mp {1\over 2.5!}
\Gamma^{\lambda \nu \rho \alpha \beta}F^{(5)}_{\lambda \nu \rho \alpha
  \beta}\Big]\Gamma_{\mu}\epsilon_{\mp}\,,
\label{gravitino}
\end{eqnarray}
where $\mu, \nu ,\rho, \lambda$ are ten dimensional space-time
indices, and hated indices refer to the Lorentz frame.
Note that for the background (\ref{I-ppwaveg}) - (\ref{I-ppwaveH}), the vanishing of the
above supersymmetry variations leads to the Killing spinors,
\begin{eqnarray}
\epsilon^{(bg)}_\pm = e^{\pm\frac{\mu}{\sqrt{2}}x^+\left(\Gamma^{\hat 1\hat 2}\cos \alpha + \Gamma^{\hat 3\hat4}
\sin \alpha\right)}\epsilon^{(0)}_\pm \,,\qquad \Gamma^{\hat +} \epsilon^{(0)}_\pm=0\,,
\label{ks-bg}
\end{eqnarray}
where $\epsilon^{(0)}_\pm$ are constant spinors. Now let us solve the above dilatino and gravitino
varioations for the (D1-D5)-brane solution presented in (\ref{d1-d5}). First the dilatino variation
gives the following condition on the spinors.
\begin{eqnarray}
\Gamma^{\hat i}\epsilon_{\pm} - \Gamma^{\hat + \hat - \hat i}\epsilon_{\mp} = 0,
\label{d1-cond}
\end{eqnarray}
\begin{eqnarray}
\Gamma^{\hat i}\epsilon_{\pm} + \frac{1}{3!}\epsilon_{\hat i\hat j\hat k \hat l} \Gamma^{\hat j \hat k \hat l} \epsilon_{\mp} = 0, \label{d5-cond}\end{eqnarray}
\begin{eqnarray}
\left(\Gamma^{\hat + \hat 1 \hat 2} + \tan \alpha \Gamma^{\hat + \hat 3 \hat 4}\right)\epsilon_{\mp} = 0.
\end{eqnarray}
The first two are the usual D1-brane and D5-brane supersymmetry condition even in the flat space. The third condition
is related to the property of this particular PP-wave spacetime. We will come back to this after solving
the gravitino variations more carefully.
On the other hand, solving gravitino variations gives the following conditions on the spinors
\begin{eqnarray}
\delta\psi^{\pm}_{+} &\equiv& \partial_{+}\epsilon_{\pm} \mp \frac{\mu}{\sqrt{2}}(f_1 f_5)^{-1/2}
\left(\Gamma^{\hat 1\hat 2}\cos \alpha + \Gamma^{\hat 3\hat 4}\sin\alpha \right)\epsilon_{\pm} = 0, \>\>\> \delta\psi^{\pm}_{-} \equiv \partial_{-}\epsilon_{\pm} = 0 \nonumber \\
\delta\psi^{\pm}_{i} &\equiv& = \partial_{+}\epsilon_{\pm} + \frac{1}{8}\left[\frac{f_{1,i}}{f_1} + \frac{f_{5,i}}{f_5}\right]\epsilon_{\pm} = 0, \>\>\>\> \delta\psi^{\pm}_{a} \equiv \partial_{a}\epsilon_{\pm}
= 0 \ ,
\end{eqnarray}
where in writing down the above variations, we have made use of the condition
\begin{eqnarray}
\Gamma^{\hat +}\epsilon_{\pm} = 0,
\label{light-cone}
\end{eqnarray}
and the D-brane supersymmetry conditions written in (\ref{d1-cond}) and (\ref{d5-cond}). In
what follows, we will discuss the supersymmetry for this brane solution for the parameter
$\sin^2 \alpha = \cos^2 \alpha = 1/2 $, as it is more evident. One can check that
imposing conditions
(\ref{light-cone}), (\ref{d1-cond}), and (\ref{d5-cond}) together with
\begin{eqnarray}
\left(1- \Gamma^{\hat 1 \hat 2\hat 3\hat 4}\right)\epsilon_{\pm} = 0 \ ,
\end{eqnarray}
all the variations are satisfied. Hence the (D1-D5)-brane solution presented above satisfies $1/8$
of the total spacetime supersymmetries, as there are only three independent conditions to be
imposed. Similar supersymmetry analysis for intersecting branes
with constant NS-NS flux has been presented in \cite{Biswas:2002yz}. The
worldsheet superymmetries can be analyzed by looking at \cite{Michishita:2002jp}. Hence we
skip the details here.

\section{Conclusions} In this paper we have found out a PP-wave background with constant three
form flux by taking a Penrose limit of a 1+1 dimensional orthogonal intersection of two stack of NS5-brans
in supergravities. This background is shown to be 1/2 superymmetric. We further find out an intersecting
(D1-D5)-brane solution in this background and study its spacetime superymmetry properties by solving the
dilatino and gravitino variations explicitly.  We have shown that unlike the PP-wave background coming
from the near horizon Penrose limit of a single stack of NS5-branes, here the nonperturbative spectrum is similar to the flat space counterpart.

\end{document}